\documentstyle[editedvolume,psfig]{crckapb}

\def\cl{\centerline}

\def\bs{\bigskip}

\def\ea{et al.\,}
\def\eg{{\it e.g.\,}}
\def\be{\begin{equation}}
\def\ee{\end{equation}}
\def\rel{relativistic\,\,}
\def\nrel{nonrelativistic\,\,}

\begin{opening}
\title{Cosmology with the S-Z Effect}

\author{Yoel Rephaeli}

\institute{School of Physics \& Astronomy, Tel Aviv University\\ 
Tel Aviv 69978, Israel\\Center for Astrophysics and Space 
Sciences\\University  of California, San Diego\\La Jolla, 
CA\,92093-0424}
\end{opening}

\begin{document}

\runningtitle{THE S-Z EFFECT}

\begin{abstract}

Extensive recent work on the Sunyaev-Zeldovich (S-Z) effect reflects 
major progress in observational capabilities of interferometric arrays, 
the improved quality of multi-frequency measurements with upcoming 
ground-based and stratospheric bolometer arrays, and the intense 
theoretical and experimental work on the small scale structure of the 
cosmic microwave background (CMB) radiation. I briefly describe the 
effect and discuss its significance as a major cosmological probe. 
Recent results for the gas mass fraction in clusters and the Hubble 
constant (largely from measurements with the BIMA and OVRO 
interferometric arrays) are discussed. Also reviewed are results from 
the first determination of the CMB temperature at the redshifts of two 
clusters (from measurements with the MITO and SuZIE experiments), and 
recent work on the CMB anisotropy due to the S-Z effect. 

\end{abstract}

\section*{Introduction}
The Sunyaev-Zeldovich (S-Z) effect is a small change in the 
intensity of the CMB that is caused by Compton scattering 
as the radiation passes through clusters of galaxies. Energy is 
transferred from the hot intracluster (IC) gas to the radiation and a 
fraction of the photons move from the Rayleigh-Jeans (R-J) to 
the Wien side of the (Planck) spectrum; the radiation is said to be 
Comptonized. A quantitative description of the effect was given 
by Zeldovich \& Sunyaev (1969) and Sunyaev \& Zeldovich (1972), who 
realized very early on the cosmological significance of this characteristic 
spectral signature. 

First measurements of the effect in a small number of clusters were made 
long ago, but these were mostly low (signal) quality results. S-Z 
observations significantly improved when interferometric arrays equipped 
with high sensitivity receivers began to be used. The current 
high-quality S-Z database, consisting largely of $\sim 30$ GHz 
measurements with the 
BIMA and OVRO interferometric arrays (\eg, Carlstrom \ea 1996, Grego \ea 2000, 
Reese \ea 2000), now includes some 60 clusters. At present, the S-Z 
cosmological results (Carlstrom \ea 2001) are based almost entirely on 
these (single frequency) measurements of a sample of moderately distant 
clusters. 

This brief review highlights recent theoretical and observational results. 
The reviews by Rephaeli (1995a) and Birkinshaw (1999) provide a more 
comprehensive discussion of most of the essential theoretical and 
observational aspects of the effect. Results from the extensive program of 
observations with interferometric arrays were reviewed recently by 
Carlstrom \ea (2001).

\section*{The Effect}

Accurate description of the change of CMB intensity due to interaction with 
fast moving electrons in clusters necessitates use of the exact frequency 
re-distribution function in Compton scattering and a relativistic 
calculation. The original description of the effect by Sunyaev \& 
Zeldovich (1972) is based on a solution to the Kompaneets equation, a 
\nrel diffusion approximation to the exact kinetic equation for the 
scattering. The \nrel calculation yields a simple expression for the 
change of CMB intensity induced by scattering of the CMB by electrons 
with {\it thermal} velocity distribution 
\be
\Delta I_{t} = i_{o} y g_{0}(x) \;,
\ee
where $i_{o}=2(kT)^3 /(hc)^2$, and $T$ is the CMB temperature. All the 
dependence on the properties of the cluster is in the Comptonization 
parameter, 
\be
y=\int(kT_e/mc^2) n \sigma_T dl \:,
\ee
an integral over the electron density ($n$) and temperature ($T_e$); 
$\sigma_T$ is the Thomson cross section. The spectral distribution 
is given in terms of the non-dimensional frequency $x \equiv h\nu/kT$,
\be
g_{0}(x)={x^4e^x \over (e^x -1)^2} \left[{x (e^x +1)\over e^x -1}-4 
\right] ;
\ee
$g_{0}(x)$ is negative for $x < 3.83$ and positive at larger values of this 
critical frequency,  $\sim 217$ GHz. The magnitude of the relative temperature 
change due to the thermal effect is $\Delta T_t/T = -2y$ in the R-J region, 
with $y \sim 10^{-4}$ along a line of sight (los) through the center of a 
rich cluster.  

Likely motion of the cluster in the CMB frame induces a {\it kinematic} 
(Doppler) S-Z component,
\be
\Delta I_k = - i_{o} h_{0}(x) \beta_{c} \tau \;, \;\;\;\;
h_{0}(x) = {x^{4}e^{x}\over (e^x -1)^2} \; ,
\ee
where $\beta_{c} = v_r/c$, with $v_r$ the line of sight component of the 
cluster peculiar velocity, and $\tau$ is the Thomson optical depth of the 
cluster. The associated temperature change of this component is 
$\Delta T_k/T= - \beta_{c}\tau$ (Sunyaev \& Zeldovich 1980). 

The above \nrel expressions for the two components of the S-Z effect 
are sufficiently accurate only at low frequencies and low gas temperatures. 
An exact \rel description of the effect is required for its use as a 
precise cosmological probe. Since electron velocities in the IC gas are 
high, the relative photon energy change in the scattering is large enough 
to require a \rel calculation. Using the exact probability distribution 
in Compton scattering, and the relativistically correct form of the 
electron Maxwellian velocity distribution, an improved calculation of 
$\Delta I_t$ was performed in the limit of small $\tau$ (Rephaeli 1995b). 
Results of this semi-analytic calculation demonstrate that the 
relativistic spectral distribution of the intensity change is quite 
different from that derived by Sunyaev \& Zeldovich (1972). Deviations 
from their expression increase with $T_e$ and can be substantial, as can 
be seen in Figure 1. These are particularly large near the crossover 
frequency, which shifts to higher values with increasing gas temperature. 

\begin{figure}[t]
\cl{\psfig{file=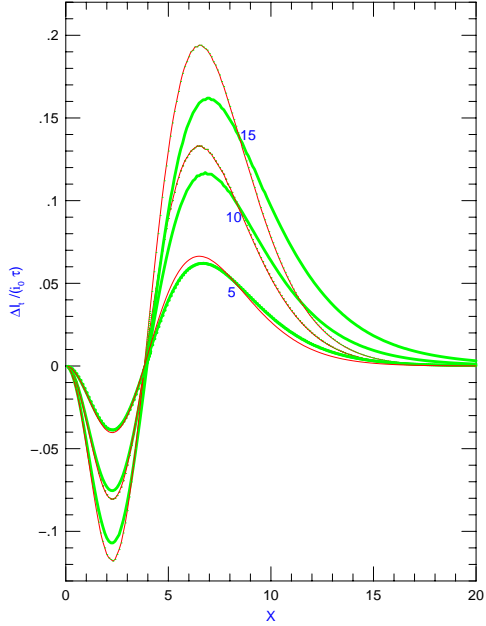,width=10cm,angle=270}}
\bs
\bs
\bs
\caption{The spectral distribution of $\Delta I_t /(i_{o} \tau)$. 
The pairs of thick (green) and thin (red) lines, labeled with 
$kT_e =$ 5, 10, and 15 keV, show the \rel and nonrelativistic 
distributions, respectively.}
\end{figure}

Relativistic generalization of the Sunyaev \& Zeldovich (1972) treatment 
has been discussed in many papers since 1995. For example, Challinor \& 
Lasenby (1998) generalized the \nrel Kompaneets equation and obtained 
analytic approximations to its solution by means of a power series in 
$\theta_{e}$ = $kT_{e}/mc^{2}$. This approach was adopted by Itoh \ea 
(1998) who improved the accuracy of the analytic approximation by 
expanding to fifth order in $\theta_{e}$. Sazonov \& Sunyaev (1998), 
and Nozawa \ea (1998a), extended the relativistic treatment also 
to the kinematic component obtaining the leading cross terms in the 
expression for the total intensity change ($\Delta I_t + \Delta I_k$) 
which depends on both $T_e$ and $v_r$. An improved analytic fit to the 
numerical solution, valid for $0.02 \leq \theta_{e} \leq 0.05$, and 
$x \leq 20$ ($\nu \leq 1130$ GHz), was given by Nozawa \ea (2000). 
In all these treatments only terms linear in $\tau$ were calculated. 
Since in some rich clusters $\tau \sim 0.02-0.03$, the approximate 
analytic expansion to fifth order in $\theta_{e}$ necessitates also 
the inclusion of multiple scatterings, of order $\tau^2$. Such terms 
were calculated by Itoh \ea (2000), and Shimon \& Rephaeli (2002). 
The relativistically generalized expression for the total intensity 
change, $\Delta I = \Delta I_t + \Delta I_k$, can be written 
(to first order in $\tau$) in the form:
\be
\Delta I =  i_{o} h_{0}(x) 
\int d\tau \biggr[\theta g_{1}(x) - \beta_{c} + R(x, \theta, \beta_{c}) 
\biggl ] \, , 
\ee
where $g_{1}(x) =  x (e^{x} +1)/( e^{x} -1) - 4$. The first two 
terms are just the \nrel expressions for $\Delta I_t$ and 
$\Delta I_k$, respectively, and the function $R(x, \theta, \beta_{c})$ 
includes the additional thermal and kinematic terms obtained in a \rel 
calculation. This function can be written in the form 
\begin{eqnarray}
R(x, \theta, \beta_{c}) & \simeq & \theta^{2} \biggr[g_{2}(x)+\theta
g_{3}(x) + \theta^{2}g_{4}(x) + \theta^{3}g_{5}(x) \biggl] \\
&  & - \beta_{c}\theta \biggr[ h_{1}(x) + \theta h_{2}(x) \biggl] \mbox{} 
+ \beta_{c}^{2}\biggr[1 + \theta h_{3}(x) \biggl] \nonumber \,, 
\end{eqnarray}
that includes terms to orders $\theta ^5$ and $\beta_{c}^{2} \theta$. 
More details on the calculation, and the definitions of the functions 
$g_{i}$ and $h_{j}$ ($2 \leq i \leq 5$ and $1 \leq j \leq 3$), can 
be found in Shimon \& Rephaeli (2002). 

Measurements with the interferometric BIMA and OVRO arrays have been 
made at (relatively) low frequencies ($\sim 30$ GHz, or $x \sim 0.5$) 
for which differences between the \nrel and the above more exact 
expressions are small. But measurements of a few clusters have also 
been made at much higher frequencies (\eg, up to $x \sim 6.2$ in the 
case of the MITO telescope, and even higher for some of the 
upcoming S-Z projects). Use of the relativistically exact expressions 
for $\Delta I_t$ and $\Delta I_k$ is clearly necessary at high 
frequencies ($\nu >> 30$ GHz). This is especially so when the effect 
is used for the purpose of determining precise values of cosmological 
parameters. Also, since the ability to determine peculiar velocities of 
clusters depends very much on measurements very close to the crossover 
frequency whose dependence on $T_e$ is approximately given by 
$x_0 \simeq 3.830016(1+1.20594\theta +2.07826\theta^{2}-80.74807\theta^{3})$ 
(Shimon \& Rephaeli 2002). Note that because high precision S-Z work 
entails use of X-ray derived gas parameters, similarly accurate 
expressions for the X-ray bremsstrahlung emissivity have to be used 
(Rephaeli \& Yankovitch 1997). In the latter paper first order 
relativistic corrections to the velocity distribution and 
electron-electron bremsstrahlung were taken into account in 
correcting values of the Hubble constant, $H_0$, that were 
previously derived using the \nrel expression for the emissivity 
(see also Hughes \& Birkinshaw 1998, Nozawa \ea 1998b).

We briefly note that there are various ways by which the scattering of the 
CMB in clusters can (linearly) polarize the radiation, generally at levels 
which are below 1 $\mu$K (Sazonov \& Sunyaev 1999). First, polarization is induced due to the CMB quadrupole 
component, at a maximal level of $\sim 0.1 (\tau/0.02)$ $\mu$K. 
When the cluster has a finite velocity component transverse to the line 
of sight, $v_{\perp}$, there are two contributions to the polarization 
that are $\propto (v_{\perp}/c)^2\tau$, and $\propto (v_{\perp}/c)\tau^2$.
The radiation will also be polarized when scattered in a cluster with 
aspherical electron distribution. 

The great property of the S-Z effect which makes it a uniquely 
important cosmological probe is its (essentially) redshift independence. 
Measurements of the effect yield directly the integrated pressure of 
the hot IC gas, and thereby also the total mass of the cluster. The 
cluster velocity along the los can be deduced from measurement of 
the effect close to the corossover frequency. From S-Z and X-ray 
measurements (the angular diameter distance, and therefore) $H_0$ can 
be determined. This method to measure $H_0$ has clear advantages over 
the traditional galactic distance ladder method. It is also possible to 
determine the density parameter, $\Omega$, from the Hubble diagram (or 
from the redshift dependence of the gas mass fraction) when a sufficiently 
precise database on a large sample of clusters is available. The 
feasibility of detecting clusters at large redshifts strongly motivates 
performing number counts through cluster surveys in order to 
characterize the population and its cosmological 
evolution. Also, mapping the CMB anisotropy induced by clusters can 
yield important information on the cluster mass function, cluster 
properties and evolutio of the population. Finally, the anisotropy 
and redshift evolution of the CMB temperature, $T(z)$, can be tested 
and verified through multi-frequency measurements of the effect in 
clusters at different sky directions and redshifts. A brief summary 
of current results is given below.

\section*{Recent Measurements}

Sensitive ground based measurements of the thermal and kinematic 
components of the S-Z effect with single dish telescopes face the 
major challenge of accounting for atmospheric emission. The significance 
of astrophysical sources of confusion -- such as emission from Galactic 
dust, cluster radio sources, and CMB anisotropy -- varies greatly with 
the frequency of observation and telescope 
beam size. Over the last decade observational S-Z work has been carried 
out mostly with interferometric arrays. These have major advantages over 
single dish telescopes, including sensitivity to specific angular scales 
and to signals which are correlated between array elements, insensitivity 
to changes in atmospheric emission, and high angular resolution that enables 
nearly optimal subtraction of signals from point sources. The improved 
sensitivity of radio receivers made it feasible (mainly through the use 
of low-noise HEMT amplifiers) to image the effect in moderately distant 
clusters, first with the Ryle telescope (Jones \ea 1993), and then mostly 
with the BIMA and OVRO arrays (Carlstrom \ea 1996, 2001). The extensive 
program of S-Z observations with BIMA and OVRO has resulted in high 
quality measurements (at frequencies near $\sim 30$ GHz) of about 60 
clusters in the redshift range $0.17 < z <0.89$. An example of the 
higher sensitivity and resolution of interferometric images over the 
traditional one dimensional drift scans, the BIMA image of A2163 
(Carlstrom \ea 2001) is shown in Figure 2 above a profile of the 
effect in the same cluster from measurements with the small SuZIE 
array (Holzapfel \ea 1997a). The BIMA contour plot is superposed 
on a ROSAT X-ray image (false color) of the cluster; as 
expected, the cluster X-ray size appears smaller than its S-Z size.
                              
\begin{figure}
\cl{\psfig{file=fig2_1.ps,width=8cm,angle=0}}
\vspace{0.2in}
\cl{\psfig{file=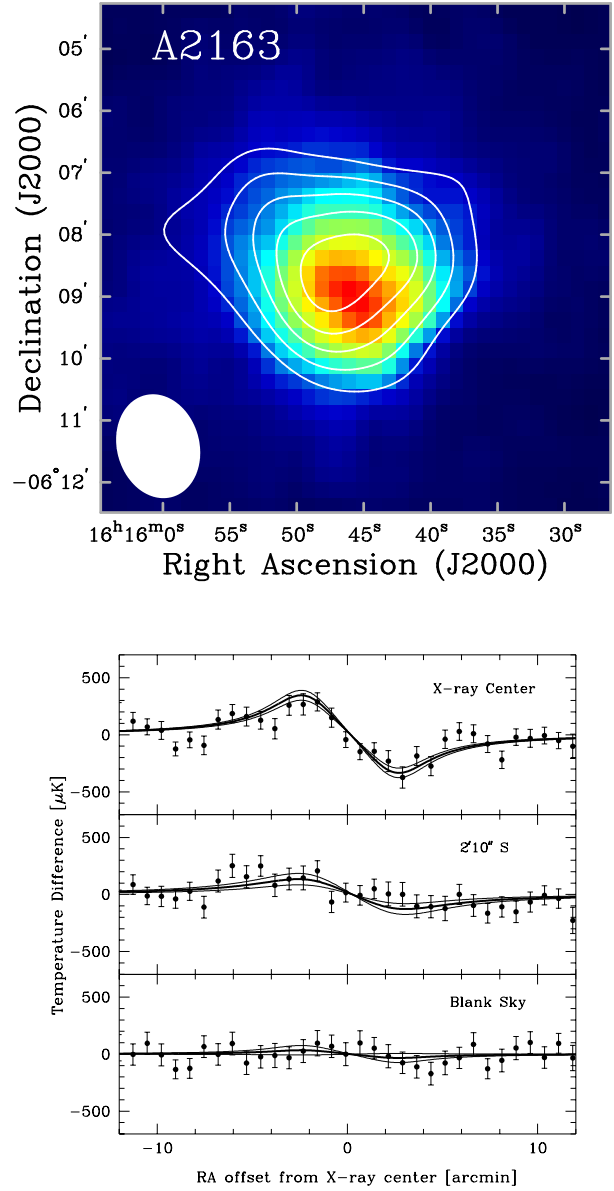,width=8cm}}%,angle=270}}
\vspace{0.2in}
\caption{S-Z and X-ray views of the cluster A2163 (from Carlstrom \ea 
2001). The upper frame shows the contour image (with the ellipse 
indicating the FWHM beam size) obtained from interferometric BIMA 
measurements (at 28.5 GHz), superposed on a (false color) ROSAT X-ray 
image. The S-Z profile from co-added drift-scan measurements (at a 
central frequency of 142 GHz) with the SuZIE array (Holzapfel \ea 1997a) 
across the center, $\sim 2.2'$ to the South, and across a blank sky 
region, are shown in the lower frame; the lines show predicted S-Z 
profiles.} 
\end{figure}  

The array configurations of BIMA and OVRO telescope systems used in the 
(above mentioned) measurements are not suitable for S-Z observations 
of nearby clusters. A more optimal system for imaging the effect over 
larger angular scales is the CBI -- an interferometric array of 13 small 
(0.9m) dishes, with spatial resolution in the $3'-10'$, operating in the 
26-36 GHz spectral range. Work with the CBI began at the Atacama desert 
(Chile), and has already resulted (Udomprasert \ea 2001) in measurements 
of the effect in 9 clusters.

There are major advantages in measuring the effect at several high 
($\nu > 100$ GHz) frequencies (close to and on the Wien side) where its 
characteristic spectral shape can be used as a powerful discriminant to 
control observational and systematic uncertainties. Multi-frequency 
measurements of the effect also enhance its scope as a cosmological 
probe; an example for this will be shown in the next section.
High frequency observations were made with the SuZIE array, the PRONAOS 
and MITO telescopes, and the Diabolo bolometer. Three moderately distant 
clusters were measured with the SuZIE $2 \times 3$ element array: A1689 
\& A2163 (Holzapfel \ea 1997a, 1997b; the S-Z profile across A2163 is 
shown in Figure 2), and A1835, which was observed at three spectral bands 
centered on 145, 221, 279 GHz (Mauskopf \ea 2000). Work with SuZIE 
continued with measurements of at 6 more clusters (S. Church, private 
communication).

PRONAOS, a stratospheric 2m telescope, measured the effect in A2163 at 
four broad spectral bands in the combined range of 285-1765 GHz (Lamarre 
\ea 1998). This seems to have been the first detection of the effect by a 
balloon-borne experiment. The MITO 2.6m telescope (located in Testa Grigia 
in the Italian Alps) originally had only a single photometer with 
a large $\sim 17'$ beam and four high-frequency bands. Even so it 
was possible to measure the effect in the Coma cluster (De Petris \ea 
2002), in spite of the fact that the fluctuating atmospheric emission 
dominated the signals. The sample of clusters in which the S-Z effect was 
measured includes the distant (z=0.45) cluster RXJ 1347 with the largest 
deduced Comptonization parameter, $y = 1.2\times 10^{-3}$ (Pointecouteau 
\ea 1999), one of 5 clusters observed 
%(Desert \ea 1998) 
with the Diabolo bolometer used at the IRAM 30m radio telescope. 

\section*{Cluster and Cosmological Parameters}

The well understood nature of the S-Z effect and its redshift independence 
have always provided strong motivation for using it as a cosmological 
probe. Availability of the high quality interferometric S-Z data obtained 
with BIMA and OVRO has significantly improved the scientific yield from 
results of S-Z and X-ray measurements. The following is only brief update 
on current results; discussion of the basic methodologies can be found in 
the reviews by Rephaeli (1995a), Birkinshaw (1999), and Carlstrom \ea (2001).

\subsection*{Cluster Quantities}

Density and temperature profiles of IC gas can be determined from X-ray 
and S-Z measurements. Spatially resolved S-Z measurements can, in 
principle, yield these distributions out to larger radii due to 
the linear dependence of $\Delta I_t$ on $n$, as compared to the 
$n^2$ dependence of the thermal X-ray brightness profile. The 
equation of hydrostatic equilibrium can then be used to find the 
cluster total mass, $M(r)$, and therefore also the 
gas mass fraction at a given radius, $r$. Grego \ea (2001) determined 
gas mass fractions in 18 clusters from interferometric BIMA and 
OVRO measurements. They assumed isothermal gas with the familiar density 
profile, $(1 + r^2/r_c^2)^{-3\beta/2}$. The core radius, $r_c$, and 
$\beta$ were determined from the S-Z data; only the temperature was 
taken from results of spectral X-ray measurements. This analysis 
yielded the gas mass fraction in the region were the S-Z data 
are sufficiently sensitive (typically 1'). Scaling relations from 
numerical simulations were then used to extrapolate the gas mass 
fraction to a distance where the cluster mean density is 500 times 
the critical density (at the appropriate redshift), since it is 
presumed that at this distance the cluster gas mass fraction well 
samples the universal baryon fraction (Evrard 1997). In the currently 
popular open and flat, $\Lambda$-dominated CDM models, mean values in 
the range $(0.06-0.09)h^{-1}$ (where $h$ is the value of $H_0$ in units 
of $100$ km s$^{-1}$ Mpc$^{-1}$) were obtained for the gas mass fraction 
(Carlstrom \ea 2001). Use of more realistic temperature profiles that can 
now be measured by XMM and {\it Chandra} will reduce the substantial 
modeling uncertainties in these mass estimates. 

To measure cluster radial velocities from the kinematic component 
of the S-Z effect observations have to be made in a narrow spectral band 
near the crossover frequency, where the thermal effect vanishes while the 
kinematic effect (which is usually swamped by the much larger thermal 
component) is maximal (Rephaeli \& Lahav 1991). This necessitates 
knowledge of the exact spectral shape of the thermal component near this 
frequency, and control of significant systematic uncertainties (such 
as due to the primary CMB anisotropy). SuZIE is the first experiment 
with a spectral band centered on the crossover frequency. Measurements 
of the clusters A1689 and A2163 (Holzapfel \ea 1997b) and A1835 
(Mauskopf \ea 2000) yielded substantially uncertain results for $v_r$ 
($170^{+815}_{-630}$, $490^{+1370}_{-880}$, and $500 \pm 1000$ km s$^{-1}$, 
respectively). Balloon-borne measurements of the effect with PRONAOS 
have also resulted in a statistically insignificant value for the 
peculiar velocity of A2163 (Lamarre \ea 1998). 

\subsection*{$H_0$ and $\Omega$}

The Hubble constant and the contributions to the cosmological density 
parameter -- that of matter, $\Omega_{M}$, and the cosmological constant, 
$\Omega_{\Lambda}$ -- can be determined from a plot of the angular 
diameter distance, $d_A$, vs. redshift (Hubble diagram). The ability to 
infer $d_A$ from S-Z and X-ray observations is essentially based on the 
different density dependences of Comptonization and thermal bremsstrahlung.
Specifically, $d_A$ can be deduced from $\Delta I_t$, the X-ray surface 
brightness, and their spatial profiles. It is (at least currently) 
unrealistic to obtain a usefully precise value of $H_0$ from measurements 
of any single cluster, due mostly to large systematic uncertainties. 
Modeling and other systematic errors can be significantly reduced by 
averaging over values of $H_0$ from measurements of a large number of 
clusters. First, $H_0$ can be determined from a sufficiently large sample 
of nearby clusters (whose angular diameter distances are little affected 
by the geometry of the universe). An initial sample of eight measured 
values yielded $H_0 \simeq 58 \pm 6$ km s$^{-1}$ Mpc$^{-1}$ (Rephaeli 
1995b). A similar mean value ($60$ km s$^{-1}$ Mpc$^{-1}$) was deduced by 
Birkinshaw (1999) based on a somewhat updated data set. The 
interferometric BIMA and OVRO S-Z survey provides the first relatively 
uniform dataset for the determination of $H_0$. Since the set 
of 33 available values (from single dish as well as interferometric 
measurements) for $d_A$ includes measurements of clusters up to $z \sim 
0.55$, the deduced mean value depends on the cosmological model: 
$H_0 = 63 \pm 3$ km s$^{-1}$ Mpc$^{-1}$ for a flat model with 
$\Omega_M = 0.3$ and $\Omega_{\Lambda} = 0.7$, and $H_0 = 60 \pm 3$ 
km s$^{-1}$ Mpc$^{-1}$ for an open model with $\Omega_M = 0.3$ (Carlstrom 
\ea 2001). While this overall $\sim 5\%$ ($1\sigma$) observational error 
is small, the estimated systematic uncertainty is $\sim 30\%$.

In principle, the dependence of the cluster gas mass fraction on the 
angular diameter distance provides a way to constrain the matter and 
cosmological constant density parameters. In a flat model, 
$\Omega_{M} + \Omega_{\Lambda} = 1$, the fit to the interferometric 
data yields $\Omega_{M} \sim 0.25$ (Carlstrom \ea 2001). 

\subsection*{CMB Temperature}

Measurements with the COBE/FIRAS experiment have shown that the CMB 
spectrum is a precise Planckian with $T_{0} = 2.725 \pm 0.002 \ $K at 
the current epoch (Mather \ea 1999). In the standard cosmological model, 
$T(z) = T_{0} (1+z)$, a fundamental relation which has not yet been 
observationally confirmed to the desired extent. Cosmological 
models with a purely blackbody spectrum but with a different $T(z)$ 
dependence than in the standard model are -- formally, at least -- 
unconstrained by the FIRAS measurements. Also unconstrained are models 
with spectral distortions that are now negligible, but may have 
been appreciable in the past. Thus far $T(z)$ has been determined 
mainly from measurements of microwave transitions in interstellar 
clouds in which atoms and molecules are excited by the CMB; for more 
on this see the review by LoSecco \ea (2001). The temperature has been 
determined in the Galaxy, as well as in clouds at redshifts up to 
$z \sim 3$. Results are, however, substantially uncertain due to the 
poorly known physical conditions in the absorbing clouds.

The use of the thermal S-Z effect to measure $T(z)$ has been 
suggested long ago. Fabbri, Melchiorri \& Natale (1978) proposed 
spectral mapping of the effect near the crossover frequency whose 
value depends on $T(z)$. This is very challenging given the dominating 
confusing signals -- such as due to the primary CMB anisotropy and the 
kinematic S-Z component -- and the fact that in the \rel treatment (of 
the effect) the crossover frequency depends on gas temperature, whose 
value has to be known very precisely if this method is to yield useful 
results for $T(z)$. Rephaeli (1980) proposed a more feasible method 
which exploits the steep frequency dependence of $\Delta I_t$ on the 
Wien side, and the weak dependence of {\it ratios} of the intensity 
change at different frequencies on the properties of the cluster. 
Formally, in the \nrel limit such a ratio is completely independent 
of the Comptonization parameter. Most of the dependence on the 
cluster parameters still drops out also in the exact \rel description, 
but a weak dependence remains on the gas temperature. When unknown, 
the cluster velocity introduces a small systematic uncertainty (which 
is, however, much smaller than in the case of measurements near the 
crossover frequency). Thus, S-Z measurements have the potential of 
yielding more precise values of $T(z)$ than can be obtained from 
ratios of atomic and molecular lines. 

With the availability of spectral measurements of the S-Z effect, the 
method of Rephaeli (1980) has recently been employed to measure $T(z)$ 
in the Coma and A2163 clusters (Battistelli 2002). Spectral measurements 
of Coma ($z= 0.0231 \pm 0.0017$) included observations at central 
frequencies $143$, $214$, and $272$ GHz (each with a $\sim 30$ GHz 
bandwidth), made with with the MITO telescope (De Petris \ea 2002), 
and the low frequency $32$ GHz (with a $\sim 13$ GHz bandwidth) 
OVRO measurement (Herbig \ea 1995). Measurements of A2163 ($z=0.203 
\pm 0.002$) were made with the SuZIE array (Holzapfel \ea 1997a) at 
$142$, $217$, and $268$ GHz (with $\sim 13-26$ GHz bandwidths), and 
a combined BIMA and OVRO measurement at $\sim 30$ GHz (LaRoque \ea 2002). 
The measurements yield three independent intensity ratios for each 
cluster; all combinations of these ratios were compared to the 
theoretically predicted values. The latter were calculated by 
performing integrations of the \rel expression for $\Delta I_t$ 
over the spectral bands of the MITO photometer, and of the other 
receivers when available. Fits of the measured ratios to the predicted 
values were performed, yielding best fit values for the CMB temperature 
at the redshifts of the two clusters. (For more on the data analysis 
and estimates of the uncertainties, see Battistelli \ea 2002.)

The results for the temperatures are $T_{Coma} = 2.789^{+0.080}_{-0.065}$ 
K and $T_{A2163} = 3.377^{+0.101}_{-0.102}$ K (at 68\%
confidence). These values are consistent with those expected from 
the standard relation $T(z)=T_{0}(1+z)$. Battistelli \ea (2002) have 
also tested two alternative scaling relations that are conjectured 
in non-standard cosmologies, $T(z)= T_{0}(1+z)^{1-a}$, and  
$T(z)=T_{0}[1+(1+d)z]$ (e.g., Lima et al. 2000). They 
determined the best fit values for the two parameters to be 
$a=-0.16^{+0.34}_{-0.32}$, and $d = 0.17 \pm 0.36$ (at 95\% 
confidence). Within the large uncertainties these values 
are consistent with zero, so no significant deviation is found 
from the standard model. LoSecco \ea (2001) obtained $a=-0.05\pm 
0.13$ and $d= 0.10\pm 0.28$ (at 95\% CL) from measurements of 
microwave transitions. The two sets of results are consistent. 
Thus, the S-Z results of Battistelli \ea (2002) already provide 
the same level of precision even though the two clusters are at 
much lower redshifts than the galaxies in the observational sample 
used by LoSecco \ea (2001). With more precise multi-spectral S-Z 
measurements expected in the future, it is anticipated that the 
S-Z method will provide a preferred alternative to the atomic 
and molecular lines method.

\subsection*{S-Z Anisotropy}

The anisotropy in the CMB induced by clusters has been extensively 
explored (since it was first modeled [Rephaeli 1981] in the context of 
a simple model for IC gas evolution), reflecting increased realization 
of its significance on arcminute scales. Since this anisotropy arises 
from the scattering of the CMB in the evolving population of clusters, 
its power spectrum and cluster (S-Z) number counts can potentially yield 
important information on the properties of IC gas, the cluster mass 
function, cosmological evolution of clusters and their gaseous contents, 
as well as some of the global cosmological and large scale parameters. 
Clearly, therefore, a quantitative description of this anisotropy 
entails the added need (when compared with a calculation of the 
primary CMB anisotropy) of modeling gas properties across the evolving 
population of clusters. 

The usual approach to the calculation of the S-Z anisotropy is based 
on the Press-Schechter model for the cluster mass function, $n(M,z)$, 
the comoving density of clusters of mass M at redshift z. 
Following collapse and virialization, IC gas is presumed to have reached
hydrostatic equilibrium at the virial temperature, with a density 
distribution that is commonly assumed to have an isothermal $\beta$ 
profile. The mass function is normalized by specifying the mass variance 
on a scale of $8\,Mpc\, h^{-1}$, $\sigma_8$, a parameter that is 
determined from the observed X-ray temperature function by using a 
mass-temperature calibration. The calibration is limited to clusters 
at small redshifts. The cluster induced anisotropy has been 
studied at an increasingly greater degree of sophistication and detail
and in wide range of cosmological and dark matter models beginning about 
a decade ago (\eg, Makino \& Suto 1993, Bartlett \& Silk 1994). In 
particular, Colafrancesco \ea (1994) calculated the temperature 
anisotropy in a flat CDM model including gas evolution (based on 
results from the Einstein Medium Sensitivity Survey [EMSS]). They later 
extended the work to other cosmological models and estimated also that many 
thousands of clusters are expected to be detected during the planned
Planck survey (Colafrancesco \ea 1997). S-Z maps and power spectra can 
also be generated directly from hydrodynamical simulations (\eg, 
da Silva \ea 2000). The range of cosmological models was extended to 
include currently viable $\Lambda$CDM models (\eg, Komatsu \& Kitayama 
1999, Molnar \& Birkinshaw 2000, Cooray \ea 2000).

Reflecting the parameter intensive nature of the S-Z anisotropy (with 
the added complexity due to the need to describe the dynamical as 
well as hydrodynamical evolution of clusters), results for predicted 
power spectra and number counts span a wide range even when calculated for 
the same cosmological model. Aside from obvious reasons for some of 
the differences, such as differing choices of gas parameters and degree 
of the evolution of the gas mass fraction, results differ also due to 
inconsistent choice of parameters whose values are coupled. An example 
is a conflicting choice of the adopted value of $\sigma_8$, and a 
different mass-temperature relation than assumed in order to determine 
$\sigma_8$ itself (for more on this, see Sadeh \& Rephaeli 2002).
 
\begin{figure}[t]
\cl{\psfig{file=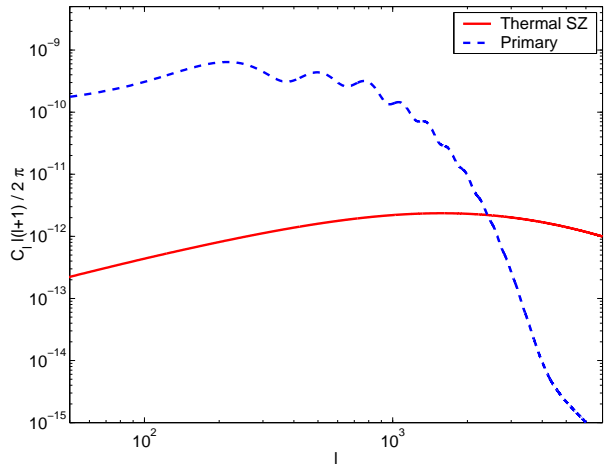,width=8cm,angle=0}}
\caption{\small{Primary and S-Z power spectra in a flat  $\Lambda$CDM model
(Sadeh \& Rephaeli 2002). The dashed line shows the primary 
anisotropy as calculated using the CMBFAST code of Seljak \& Zaldarriaga
(1996). The solid line shows the thermal S-Z power spectrum.}}
\end{figure}

The main features of the power spectrum of the thermal component of 
the S-Z anisotropy are shown in Figure 3; results are from the work of 
Sadeh \& Rephaeli (2002) who have calculated power spectra of the S-Z 
and primary anisotropies in an array of cosmological and dark matter 
models. The S-Z power spectrum, based on a Press \& Schechter mass 
function, was normalized by the observed X-ray luminosity function. 
The primary power spectrum (dashed line) was calculated using the 
CMBFAST code of Seljak \& Zaldarriaga (1996). The plots are of the 
predicted power spectrum, $C_{\ell}(\ell +1)/2\pi$ vs. multipole $\ell$, 
in a flat cosmological model with $\Omega_{\Lambda} = 0.7$, and CDM 
density parameter $\Omega_M = 0.3$. IC gas was assumed to evolve in 
a simple manner consistent with the results of the EMSS survey, as 
parametrized by Colafrancesco \ea (1994). In this model the S-Z 
anisotropy (which is dominated by the thermal effect; 
the contribution of the kinematic effect is of second 
order) is appreciable already at $\ell \sim 1500$. Clearly, the 
S-Z anisotropy has to be accounted for in the detailed modeling 
of the small scale structure of the CMB, if parameter extraction 
is to be precise. The first observational evidence for S-Z anisotropy 
could possibly be the recent CBI measurement of excess power (at an 
estimated $3.1\sigma$ level) over what is predicted in standard models 
for the primary anisotropy at multipoles $\ell = 2000$--$3500$ (Mason 
\ea 2002, Bond \ea 2002).

\section*{Prospects for the Near Future}

The quality of the scientific yield from the many S-Z images obtained 
over the last few years with interferometric arrays has made quite 
apparent the great potential of the S-Z effect as a major cosmological 
probe. This is just the beginning: Numerous S-Z projects will become 
operational in the near future, both bolometric multi-frequency arrays,
and interferometric arrays; these include the ground-based (upgraded) 
MITO telescope, the SZA, AMiBA, and the balloon-borne telescopes 
BOOST and OLIMPO. Improved sensitivity, higher spatial resolution, and 
expanded spectral range will greatly improve the quality of S-Z 
measurements. Better understanding and control of systematics will 
continue to determine the overall value of cosmological results 
from S-Z and X-ray measurements. Beacause of this, optimal results 
will likely be obtained from measurements of the effect in nearby 
($z \leq 0.1$) clusters. For example, higher quality S-Z and X-ray 
measurements of a sufficiently large sample of nearby clusters are 
expected to result in an overall uncertainty of $\sim 5\%$ in the 
value of $H_0$.

\end{document}